\documentclass[conference]{IEEEtran}

\usepackage{cite}
\usepackage{amsmath,amssymb,amsfonts}
\usepackage{algorithmic}
\usepackage{graphicx}
\usepackage{textcomp}
\usepackage{subcaption}
\usepackage{xcolor}
\def\BibTeX{{\rm B\kern-.05em{\sc i\kern-.025em b}\kern-.08em
    T\kern-.1667em\lower.7ex\hbox{E}\kern-.125emX}}
\begin{document}

\title{Towards Obfuscated Malware Detection for Low Powered IoT Devices}


\author{
    \IEEEauthorblockN{
        Daniel Park, 
        Hannah Powers, 
        Benji Prashker, 
        Leland Liu
        and B\"ulent Yener}
    \IEEEauthorblockA{Rensselaer Polytechnic Institute, Troy, NY, USA}
    \IEEEauthorblockA{\{parkd5, powerh, prashb, liul8\}@rpi.edu}
    \IEEEauthorblockA{yener@cs.rpi.edu}
}

\maketitle

\begin{abstract}
With the increased deployment of IoT and edge devices into commercial and user networks, these devices have become a new threat vector for malware authors. It is imperative to protect these devices as they become more prevalent in commercial and personal networks. However, due to their limited computational power and storage space, especially in the case of battery-powered devices, it is infeasible to deploy state-of-the-art malware detectors onto these systems. In this work, we propose using and extracting features from Markov matrices constructed from opcode traces as a low cost feature for unobfuscated and obfuscated malware detection. We empirically show that our approach maintains a high detection rate while consuming less power than similar work.
\end{abstract}


\section{Introduction}
While strides have been made in malware detection, such as in \cite{saynotooverfitting}, many solutions are not feasible in computationally or power constrained environments, such as battery-powered Internet of Things (IoT) devices. A consequence of the increased popularity of IoT devices is the addition of a large attack vector for computer networks. Because of their large numbers, weak computational capabilities, and flawed security \cite{ZhangZhi14}, they are currently a network's weakest link, as was seen in the 2017 Mirai Botnet attack \cite{Kolias17}. This makes it imperative that such devices are outfitted with malware detection tools that can be run real-time without relying upon other network devices as to (1) not be bottle-necked by network communications, (2) not rely on devices that may already be infected, and (3) avoid sending massive amounts of data to a single computation node. However, the problem is complicated due to obfuscated malware, or malware that have undergone transformations in an attempt to hide their malicious behaviour.

However, the limited computational power of these devices is currently a bottleneck and has led to increased research in communication and resource management to improve their efficiency \cite{cao19,mao17,luong18}. Further research has also been conducted specifically for battery-powered devices \cite{Oz17, Samie16, Singh17}. Unfortunately, current IoT malware detection research  ignores the energy consumption or computational power needed to deploy machine-learning based malware detection tools on such devices. It is imperative to consider power consumption in deployed security solutions such that the consumption still allows a high level of autonomy for the device \cite{Senni16}.

There have been innovative solutions to malware detection and classification, such as using convolutional and recurrent neural networks \cite{Nataraj11,raff2017}. However, these models typically cannot be deployed onto IoT devices because of their power, storage, and computational costs. N-gram approaches are also popular, however are restricted by the chosen vocabulary and size of $n$. In this work, we explore the use of Markov chains for malware detection as they have been shown to be an expressive representation of the feature space without incurring high storage costs \cite{Anderson12, Runwal2012, Shafiq2008}.

\textbf{Research Question.} Successful integration of IoT devices and edge devices onto a network depends on the security of those devices. Can we deploy malware detection models onto such devices that are robust enough to detect obfuscated malware samples?

\textbf{Our contributions.} In this paper, we propose the use of Markov matrices and Markov chains for obfuscated x86 IoT malware detection. We experimentally show that generating these Markov matrices, as well as extracting additional features from these matrices, results in significantly lower power consumption while maintaining a higher detection accuracy when compared to related work. 

\textbf{Organization.} The rest of the paper is organized as follows. In Section \ref{related-work} we cover related approaches. In Section \ref{background} we summarize current malware detection and obfuscation methods. In Section \ref{threat-model}, we discuss the threat model considered in this study. Then, in Section \ref{methodology} we introduce new features and cover existing features for malware detection. We show empirical results in Section \ref{experimental-results} and conclude in Section \ref{conclusion}.

\section{Related work} \label{related-work}
Malware detection on IoT or edge devices is not a new concept. However, to the best of our knowledge, no study considers both malware detection and resource constraints. Additionally, many studies are specific to the Android platform and use platform-specific features.

   \subsubsection{IoT malware detection}
    The area of IoT malware detection is huge and cannot be covered in this section alone. We will only be focusing on popular platform agnostic features that do not require domain knowledge. 
        
    Most IoT malware research focuses on Android malware. This is because the availability of datasets, such as Drebin, and its popularity in mobile devices \cite{Arp14}. For example, DroidMat uses dynamic analysis to collect a sequence of API calls \cite{Wu12}. Riasat et. al. uses both static and dynamic analysis to extract features such as permissions and API calls from Android application package (APK) manifest files \cite{Riasat17}. DroidSieve attempts to detect malware only using a collection of static features, such as permissions, authentication methods, and certificate information extracted from the APK \cite{Suarez17}.
        
    The above works evaluate their feature sets using Support Vector Machines, Random Forests, and e\textbf{X}treme \textbf{G}radient \textbf{B}oosting (XGBoost) \cite{Arp14,Smutz12,saynotooverfitting}. Suarez-Tangil et. al. found that gradient boosted trees provided the best classification results, and this is further validated by the results of Kaggle's MMBIG competition \cite{Suarez17, Ronen18}.

    As previously stated, the current malware detection literature assumes resource (both computational and energy) rich environments. For example, DroidSieve presented very high detection and classification accuracy, reporting between $92\%$ and $100\%$ accuracy. They claim their feature set can be inexpensively extracted. However, this is on the assumption that DroidSieve is being run on a powerful machine. The authors reported that their experiments were done on a single core of Intel® Xeon® Processor E5-2697 v3, which provides more computational resources in terms of processor frequency, threads, and cache than all currently available Intel® Quark® microcontrollers, which are designed for IoT applications.
  
   \subsubsection{Markov chains for malware detection}
   We are not the first to propose the use of Markov matrices for malware detection. We propose using the Markov matrix, or the stochastic matrix, and features extracted from it, whereas the related works in this area uses the Markov chain. The related works treat the Markov chain as a directed graph and use graph similarity techniques, such as kernels, to detect malware. Shafiq et. al. proposed Markov matrices as a more robust counterpart to $n$-grams \cite{Shafiq2008}. The proposed method used the entropy rate with respect to the binary at different locations to detect malware. This is with the assumption that malicious byte sequences embedded in benign files have higher entropy. Similarly, Runwal et. al. proposed using Markov matrices to detect embedded malware using graph similarity \cite{Runwal2012}. However, both methods consider a very specific threat model. Both previous studies focus on embedded malware datasets. Shafiq et. al. created embedded malware using an "in-house" unreleased tool named \textit{Nergal}, however do not detail how this program behaves or how it embeds the malicious code into benign code. Because of this, we assume that NERGAL appends malicious code to the benign code, patches some function call or address to point to the new code. The authors also use a benign dataset, comprising of EXE's, PDFs, ZIPs, JPGs, and PDFs, and a non-embedded malware dataset. The metamorphic quality of their malware samples comes from the embedding. This is to say that malware variants or mutations appear different because the benign file to which the malware is embedded changes drastically. This means that their experimental results confirm that Markov matrices can be used in conjunction with a similarity metric to detect 1) a sub-sequence of opcodes deviates from the rest of the binary as defined by a Markov chain and 2) if a sub-sequence of opcodes transitions are similar to known malware. Neither work completely addresses obfuscating transformations applied to the malicious code itself. For example, it is unclear whether obfuscating the malicious code itself will alter the entropy of the binary enough to evade the method proposed in \cite{Shafiq2008}.
   
   Anderson et. al. combines 6 different views of the binary to create a feature vector, a control flow graph, and a Markov chain \cite{Anderson12}. These are then used with similarity measures to generate a feature vector as input to a support vector machine. Furthermore, the above works use a linear view of the opcode trace, whereas we propose augmenting our Markov chains with control flow information. This is further explained in Section \ref{methodology}. In Section \ref{experimental-results}, we compare our proposed method to that of Anderson et. al. as it is more recent and outperforms the work of Shafiq et. al. and Runawal et. al.

\section{Background} \label{background}
In this section, we summarize the problem of malware detection and will briefly discuss a popular transformation that are used to extract features from raw binary files. Then we define obfuscation and summarize popular obfuscation transformations that are used in our evaluation in Section \ref{experimental-results}.

  \subsection{Malware detection using machine learning}
  As more malware is created, and as more data is being created and transmitted on the internet, researchers have begun to turn to machine learning as a solution. There are currently a wide range of features used in the detection and classification of malware, such as image generation \cite{Zhou17, Nataraj11} for static analysis and call graph based features for dynamic analysis \cite{Kinable10}, that have been proposed to classify new or unseen malware. However, as seen during the Microsoft Malware Classification Challenge \cite{Ronen18}, $n$-grams is the most widely used feature for classification.

  \subsubsection{$N$-grams}
  $N$-grams have been a very popular feature to use for both the classification and detection of malware. In 2009, Santos et. al. proposed using  $n$-grams as a replacement to file signatures for detecting malware using a $k$-nearest neighbor model. In doing so, they showed that machine learning and $n$-grams can successfully be used to detect malware\cite{Santos09}.
  
  However, $n$-grams are dependent on two values $n$ and $l$. $n$ is the size of the gram or sliding window and $l$ is the total number of words being considered for the gram or the size of the vocabulary. In the case of x86 instructions, one might consider a 4-gram consisting of all existing instructions. Such a 4-gram would look something like (\textit{mov, mov, eax, jz}) where the 4-tuple consists of consecutively appearing instructions. Higher values of $n$ and larger vocabularies lead to a more robust view of each sample, but incur a higher storage cost. 
  
  \subsection{Obfuscation}
  Collberg et. al. defined obfuscation as a transformation which takes a program $P$ and returns another program $P'$ with the same observable behavior, however, non-visible side effects are allowed \cite{Collberg97}. Obfuscation can be used by developers with or without malicious intent. For example, a software engineer may intentionally obfuscate their code for information property reasons or to inhibit program tampering. However, it is more popularly used by malware authors to (1) hide the malicious intent of their programs for the purpose of evasion and (2) make reverse engineering and analysis difficult for the purpose of persistence.
  
  In this section, we will review the obfuscation techniques implemented in this study by summarizing definitions laid out in \cite{Park19,You10}. The transformations that we considered are only those which create variants of themselves, affecting the sequence of opcodes in a binary. Although obfuscation techniques and research have evolved, work such as Park et. al. show that these simpler transformations are still effective against machine learning based malware detection and classification models \cite{Park19}.
  
\begin{figure}[h]
      \centering
      \includegraphics[width=\linewidth]{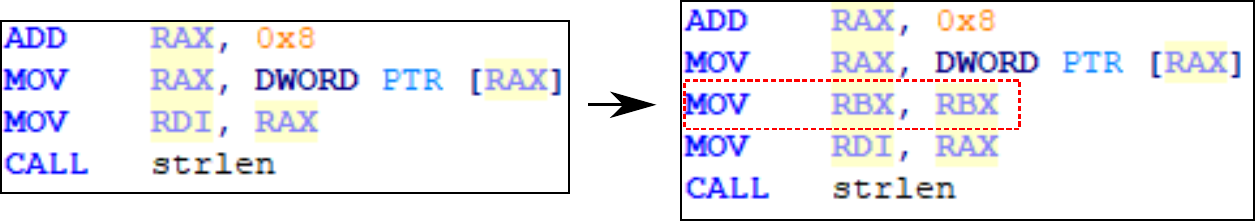}
      \caption{On the left is an x64-ELF assembly sample that takes a string from an array and calls the function \textit{strlen}. On the right is the same sample with a slightly more complicated \textit{nop} inserted. Both have the same output behavior as the original, however, the added instruction changes the samples' signature, $n$-gram, and other features.}
      \label{fig:dummycode}
  \end{figure}

  \textbf{Dead code insertion}
  The purpose of dead or dummy code insertion is to change the appearance of the binary by inserting an instruction or a sequence of instructions without changing the original program logic. The simplest method of inserting dead code is to insert a literal \textit{no operation} or a \textbf{\textit{nop}}. However, a \textit{nop} can also be achieved by selecting a semantically equivalent instruction or a sequence of instructions such as in figure \ref{fig:dummycode}.

  It is important to note that \textit{nops} are still executed and take a well-defined number of CPU clock cycles.

  \begin{table*}[ht!]
  \caption{This table lists the semantic \textit{nops} that are inserted to create executable adversarial malware examples. These assembly instructions are \textit{nops} or behave as \textit{nops} in x64.. Many more \textit{nops} and dummy code sequences exist, however, we presents results using these 22 semantic \textit{nops}.}
  \label{table:noptable}
  \centering
  \begin{tabular}{|l||l|l|l|l|}
  \hline
  \textbf{Hex} & 90                     & 6690              & 0f18000000000                    & 6689c9            \\ \hline
  \textbf{Asm} & nop                    & xchg ax, ax       & nop DWORD PTR{[}rax+0x0{]}       & mov cx, cx        \\ \hline \hline
  \textbf{Hex} & 87c9                   & 87d2              & 0f1f440000                       & 88db              \\ \hline
  \textbf{Asm} & xchg ecx, ecx          & xchg edx, edx     & nop DWORD PTR{[}rax+rax*1+0x0{]} & mov bl, bl        \\ \hline \hline
  \textbf{Hex} & 89f6                   & 88c0              & 0f1f4000                         & 5159              \\ \hline
  \textbf{Asm} & mov esi, esi           & mov al, al        & nop DWORD PTR{[}rax+0x0{]}       & push rcx, pop rcx \\ \hline \hline
  \textbf{Hex} & 0f1f00                 & 6689c0            & 6689db                           & 6687db            \\ \hline
  \textbf{Asm} & nop DWORD PTR{[}rax{]} & mov ax, ax        & mov bx, bx                       & xchg bx, bx       \\ \hline \hline
  \textbf{Hex} & 6687c9                 & 5058              & 535b                             & 83e800            \\ \hline
  \textbf{Asm} & xchg cx, cx            & push rax; pop rax & push rbx; pop rbx                & sub eax, 0x0      \\ \hline \hline
  \textbf{Hex} & 89ff                   & 87db              &                                  &                   \\ \hline
  \textbf{Asm} & mov edi, edi           & xchg ebx; ebx     &                                  &                   \\ \hline \hline

  \end{tabular}
  \end{table*}
  
  In this paper, we used a list of 22 \textit{nops} for use in dead code insertion as seen in Table \ref{table:noptable}. Many more \textit{semantic nops} can be found using the Intel 64 and IA-32 architecture processor instruction set and manual \cite{intel} or by cleverly stringing together instructions that do not effect the original program logic.
  
  \textbf{Subroutine reordering}
  Subroutine reordering alters the order in which the subroutines appear in the executable by permutation. This permutation does not alter the program logic as it does not affect its execution trace. However, unlike most of the other code obfuscation techniques, possible mutations are limited at $n!$, where $n$ is the number of subroutines to be reordered. Because subroutine reordering only moves the actual functions or modules, malware that has undergone subroutine reordering can still be detected using signature based approaches by recording the signature of each function separately.
  
  \textbf{Instruction substitution}
  Instruction substitution relies on the fact that many instructions can be executed in multiple ways. Below are two common ways to utilize instruction substitution.
  
  \textit{Static value obfuscation}
  \begin{verbatim}
                        mov eax, 0x44
    mov eax, 0x12  -->  add eax, 0x04
                        xor eax, 0x5a
  \end{verbatim}
  The sequence on the right results in the same $eax$ value, but does not immediately leak the value to an automated tool.

  \textit{Binary operators}
  A second commonly used method is replacing standard binary operators, such as addition and subtraction, with equivalent code sequences. For example, a simple addition $a = x + y$ can be substituted with
  \begin{verbatim}
            r = rand()
            a = x + r;
            a = a + y;
            a = a - r;
  \end{verbatim}
  
  \textbf{Control flow mixing}
        Control flow mixing relies heavily on instructions that alter control flow, such as $jmp$, to alter the code sequence but preserve the original behavior. The number of additional \textit{jmp} instructions increases with the number of lines or blocks that are permuted, leading to different dynamic traces of instructions. Furthermore, the actual control flow graph can be altered with the inclusion of opaque predicates, e.g., a conditional that always resolves \textit{True}. Below is an example of control flow mixing.
        \begin{verbatim}
   xor eax, eax         xor eax, eax
   mov eax, 0x45   -->  cmp eax, 0
   ret                  jnz .never_reached
                        mov eax, 0x45
                        ret
        \end{verbatim}
        
\section{Threat model}\label{threat-model}
  In this section, we describe the threat model considered in this study. The main goal of the adversary is to submit a malware sample that evades detection by a machine learning model trained to detect malware. Below, we will cover the adversary's knowledge and capabilities that can be used towards accomplishing this goal.
  
  \subsection{Adversary's knowledge}
  We will take from the literature to define the knowledge of our adversary. The adversary does not know the exact specifications of the malware detector, however does know of its existence. Furthermore, the adversary is not a "smart adversary" as defined in \cite{Yener2019}, i.e., does not have knowledge of machine learning algorithms nor have access to machine learning frameworks. However, the adversary knows of the malware detector's existence and can attempt to evade detection.
  
  \subsection{Adversary's capabilities}
  The adversary is this study has the capability to obfuscate malware samples with the following obfuscating transformations: \textit{dead code insertion}, \textit{subroutine reordering}, \textit{instruction substitution}, and \textit{control flow mixing}. These transformations were given to the adversary because of their popularity in polymorphic and metamorphic malware. The adversary can obfuscate malware ad-hoc, or randomly, with the methods previously described through 1 to 4 (randomly selected) obfuscation passes. 
  
\section{Methodology} \label{methodology}
  In this proposed method, we study using flattened Markov matrices as a feature for detection of obfuscated malware that have been transformed with dummy code insertion, subroutine reordering, instruction substitution, and control flow mixing. Additionally, we extract additional features using independent component analysis and simple graph analysis measures.
    
   We also propose the use of control-flow sensitive Markov matrices. Previous work done in malware detection using Markov matrices have use a linear disassembly of a binary to build the Markov matrix. While full dynamic analysis is infeasible in resource constrained environments, we show that simply following address jumps and function calls when building a Markov matrix increases malware detection accuracy. 
  
  \subsection{Independent component analysis}
  Independent component analysis (ICA) is a method for separating multivariate signals into its sub-components that was originally proposed as a solution to the blind source separation problem \cite{Jutten91}. However, researchers has expanded upon the original paper, showing that ICA is an effective feature extraction method for classification problems with experimental results using the MNIST dataset \cite{Ozawa01,Kwak03}.
    
  In this study, we experimented with different values to represent the number of sources for each sample. However, as we are working in a different domain the number of sources refers to the size of the feature vector after dimensionality reduction via ICA. Using only the training set, we found setting the number of "sources" to be 34 resulted in the best detection rate.
  
  \subsection{Markov matrices}
\begin{figure*}
      \centering
      \includegraphics[width=0.65\linewidth]{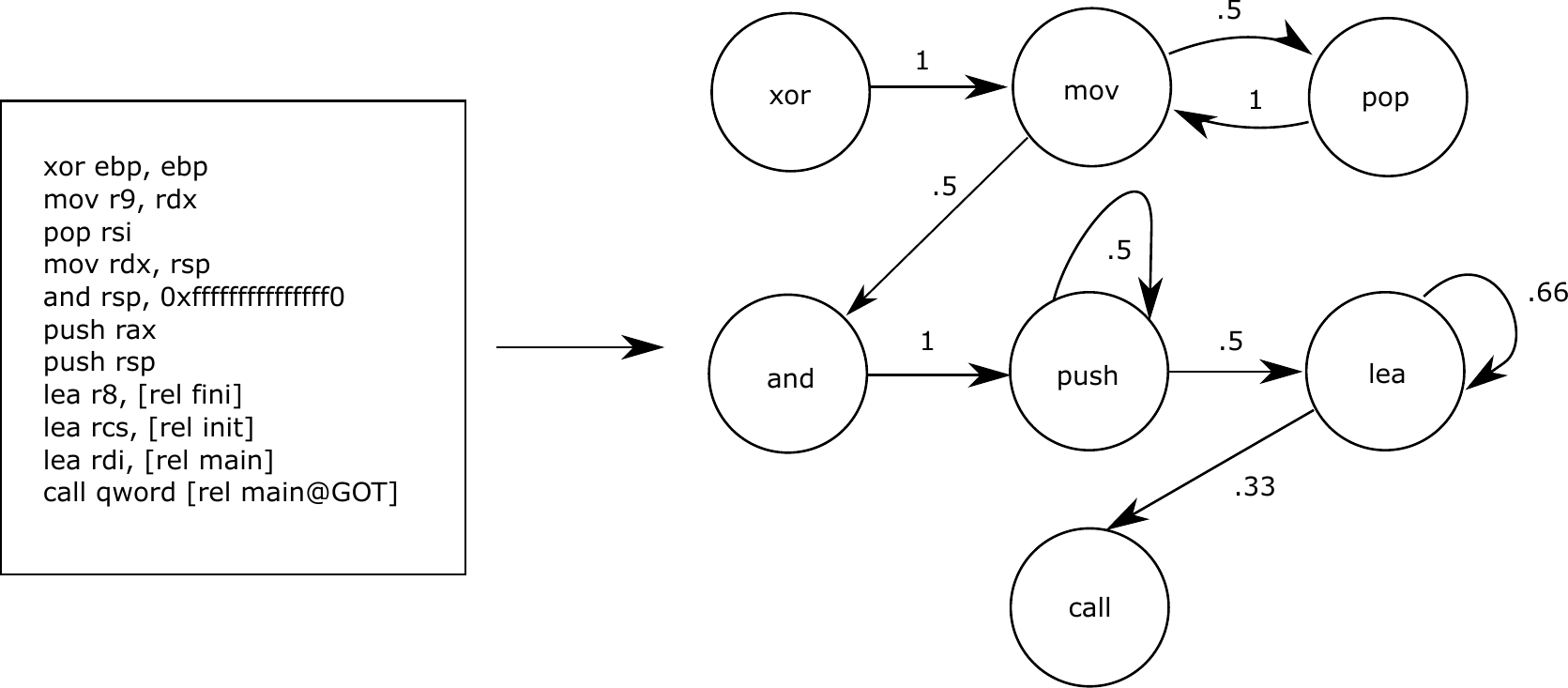}
      \caption{An example of a sequence of instructions (trace) and its corresponding Markov chain. The Markov chain's states are unique instructions found in the trace and the transition probabilities can be computed easily using bi-grams. Note that this Markov chain can be treated as a directed graph where each state is a node and the transitions are weighted edges. The Markov matrix is this graph's adjacency matrix.}
      \label{fig:markov_chain}
  \end{figure*}

  Markov chains are a sequence of states where the probability of each state occurring depends on the preceding state. In this work, we propose using Markov matrices, where the corresponding Markov chain describes a binary's opcode sequence, as a possible solution to malware detection in resource constrained IoT devices. An example of such a Markov chain and the corresponding asm code is seen in Figure \ref{fig:markov_chain}. The goal for using Markov chains, instead of $n$-grams, is to maintain a similar robust feature space without incurring a similar storage cost \cite{Shafiq2008}. Additionally, we explore the use of independent component analysis (ICA) and extraction of graph analysis metrics on Markov matrices.
  
  Simply put, a Markov matrix, or stochastic matrix, is a square matrix that describes the transition probabilities between states of a Markov chain. In this work, each state in a binary's Markov matrix represents a unique opcode or instruction found in that binary, e.g, \textit{mov}. Transition probabilities are computed by taking bi-grams $(x, y)$ from a binary's instruction trace. For example, the probability that state $y$ follows state $x$ can be computed by 
  
  $$ p_{x,y} = \frac{T_{x,y}}{T_{x,s}}$$  
  
  where $p_{x,y}$ is the probability that state $y$ is reached from state $x$, $T_{s_1, s_2}$ is the total number of $(s_1, s_2)$ bi-grams, and $s$ is any state that is preceded by $x$.
  
  \subsubsection{Control flow graph sensitivity}
  In this study, we extract Markov matrices using command line tools that come already installed on Unix systems or those that can easily be installed. To disassemble a binary to its assembly instructions, we use the command line tool \textit{objdump}. This process can be scripted to easily dump the assembly instructions located in the \textit{.txt} section of the binary. Previous methods take this assembly dump and strip all information other than the opcodes. However, we propose using the operands or arguments of control flow altering instructions (such as jmp, call, etc...) to augment the resulting Markov matrices with control flow information using rudimentary string matching techniques. For example, consider the following instruction sequence:
    \begin{verbatim}
        cmp bx, cx
        jeq istrue
        mov cx, ax
    \end{verbatim}
    
  We first extract the sequence [\textit{cmp, jeq, mov}] and generate a Markov chain from that sequence. Additionally, our proposed method takes into consideration the change in control flow to function \textit{istrue}. We augment our Markov chain with transitions from both the \textit{jmp} instruction and preceding instruction to the target address. If the target address is a function and returns, we augment the Markov chain with transitions back to the original instruction trace. All functions and return statuses can be cached for efficiency. If the target address is not a function, we augment with only a single transition to the target address's instruction. We do this with the expectation that incorporating control flow information will enable our model to better learn malicious instruction sequences, similar to how stepping into functions while manually debugging a binary possibly gives a better understanding of the binary as a whole. Using the previous example, the transition from \textit{jeq} to \textit{mov} is not indicative of the program behavior. Stepping into jump targets or called functions better represents the actual run-time behavior of the program.
  
  It is important to note that this is implemented using rudimentary string matching techniques. Furthermore, as this is a form of static analysis, it is unable to capture dynamically linked libraries. This process can be made more accurate using various dynamic analysis methods, however, would incur computational, power, and storage costs. In Section \ref{experimental-results}, we show that even this rudimentary solution increases the malware detection rate. The time complexity for generating the Markov matrix is $O(n)$ where $n$ is the number of instructions in the binary. The space complexity of the matrix is $O(V^2)$ where $V$ is the number of opcodes in our vocabulary. And the time complexity for generating the Markov chain from the Markov matrix is $O(V^2)$.

  \subsection{Graphs}
  Markov matrices can easily be converted to a graph via its corresponding Markov chain. We set the Markov matrix's states to be the nodes in the graph and the their transitions to be directed edges. From these graphs, we extracted several simple features listed below using NetworkX \cite{NetworkX}. Our goal with these graph features is to test whether it is possible to capture an expressive feature space, similar to graph matching techniques such as proposed in Anderson et. al. \cite{Anderson12}, without incurring the same high power consumption.
  
    \begin{itemize}
        \item Average Neighbor Degree
        \item Average Degree Connectivity
        \item Degree Centrality
        \item Out Degree Centrality
        \item Eigenvector Centrality
        \item Closeness Centrality
        \item Betweenness Centrality
        \item Edge Betweenness Centrality
        \item Transitivity
        \item Adjacency Spectrum
  \end{itemize}
\vspace{.8mm}
  
\section{Experimental results} \label{experimental-results}
In this section, we begin with a description of our experimental setup, including the hardware used in our experimentation. Then, we describe our malware datasets and explain how the obfuscated malware was generated. Lastly, we present our experimental results.

  \subsection{Experimental setup}
  We evaluate on two systems: (1) A desktop computer running Ubuntu 18.04 with a GeForce RTX 2070 GPU, 4-core Intel(R) Core(TM) i7-7700K CPU @ 1.15GHz, and 32GB RAM and (2) A Raspberry Pi 4 (referred to as RPi4) with a Broadcom BCM2711, 4-core Cortex-A72 (ARM v8) @ 1.5GHz and 4GB RAM. 
  
  We trained models on our desktop system and evaluated the trained models on the RPi4 to record detection accuracy and power consumption. We also report power consumption results for training models on the RPi4 to evaluate the feasibility of retraining models locally on IoT devices. We report the average detection accuracy and power consumption over 50 runs.
  
  \subsection{Datasets}\label{dataset}
  Our dataset consists of 2407 benign and 3008 malicious binaries. The benign files were collected from fresh installations of Ubuntu 18.04 and from their respective package managers.
  
  The malicious files were collected from VirusShare, an online indexed malware repository \cite{vxshare}. We used these binaries to evaluate our proposed method's effectiveness in detecting metamorphic malware. It is important to note that our evaluation in obfuscated malware detection is based on x86 Linux binaries to ensure uniformity of their opcodes. We generated Linux binaries by either applying obfuscation to the available source code or relying on state-of-the-art binary lifting and translation techniques using McSema \cite{mcsema}, Dyninst \cite{dyninst}, Obfuscator-LLVM \cite{Junod15} and custom LLVM passes. Each obfuscated binary went through 1 to 4 (randomly selected) passes of the obfuscation methods (again, randomly selected) described in Section \ref{background}.

  \begin{figure}[h]
    \centering
    \begin{subfigure}[t]{0.48\textwidth}
        \centering
        \includegraphics[width=.77\linewidth]{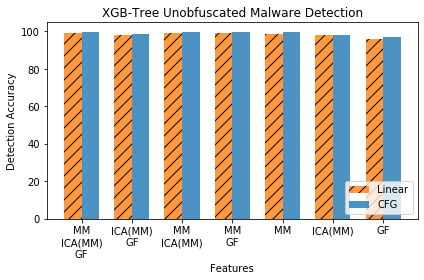}
        \caption{Unobfuscated}
    \label{fig:xgb1}
    \end{subfigure}
    ~
    \begin{subfigure}[t]{0.48\textwidth}
        \centering
        \includegraphics[width=.77\linewidth]{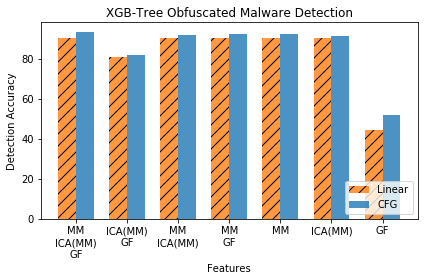}
        \caption{Obfuscated}
    \label{fig:xgb2}
    \end{subfigure}
    \caption{Unobfuscated and obfuscated malware detection accuracies using a combination of features extracted from Markov matrices (MM) using a linear and control-flow sensitive scheme. The model used for detection is an extreme gradient boosted (XGB) tree.}
    \label{fig:xgb}
  \end{figure}

  \begin{figure}[h]
    \centering
    \begin{subfigure}[t]{0.48\textwidth}
        \centering
        \includegraphics[width=.77\linewidth]{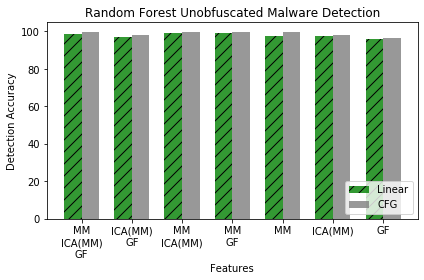}
        \caption{Unobfuscated}
    \label{fig:rf1}
    \end{subfigure}
    ~
    \begin{subfigure}[t]{0.48\textwidth}
        \centering
        \includegraphics[width=.77\linewidth]{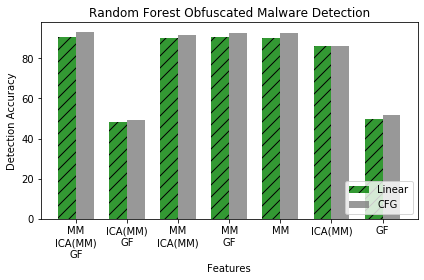}
        \caption{Obfuscated}
    \label{fig:rf2}
    \end{subfigure}
    \caption{Unobfuscated and obfuscated malware detection accuracies using a combination of features extracted from Markov matrices (MM) using a linear and control-flow sensitive scheme. The model used for detection is an random forest classifier.}
    \label{fig:rf}
  \end{figure}

  \subsection{Results}
  In Figures \ref{fig:xgb} and \ref{fig:rf}, we present results using extreme gradient boosted trees and random forest classifiers, the most popular classifiers in the malware detection domain. We show the detection rate results using the following Markov matrix extraction methods:
      \begin{itemize}
          \item \textit{Linear}: Linear Sweep
          \item \textit{CFG}: Control flow graph sensitive
      \end{itemize}
      For each of the above, we test energy consumption and detection accuracy with the following features:
      \begin{itemize}
        \item \textit{MM}: Markov matrices
        \item \textit{ICA(MM}): Independent component analysis used to extract features from the Markov matrices
        \item \textit{GF}: Graph features extracted from the Markov matrices
        \item Combinations of the above features
      \end{itemize}
  
  For each set of experiments, we test our methodology against unobfuscated and obfuscated malware. We separated our data into three dataset. The training data was kept consistent between all experiments and consisted of ~70\%  of the benign binaries and unobfuscated malware. Test set 1 (unobfuscated) consisted of the remaining ~30\% of the benign binaries and unobfuscated malware. Test set 2 (obfuscated) consisted of the same benign binaries from test set 1 and obfuscated malware. The obfuscated malware are generated using the malware samples from test set 1 as described in Section \ref{dataset}.

  We also compared our proposed features with the better performing of the two related works in Markov chains for malware detection \cite{Anderson12} in Table \ref{tab:res} and show the false positive and false negative rates in Table \ref{tab:fpn}, where positive means malicious. The false positive rate is consistent between the obfuscated and unobfuscated experiments because the benign code samples are not altered, thus their classification is consistent.
  
    \begin{table}[ht]
    \caption{Detection Accuracy Comparison against Related Work}
    \label{tab:res}
    \vspace{1mm}
    \centering
    \begin{tabular}{c l l}
    Comparison  \ \             & \textbf{Unobfuscated} \ \ & \textbf{Obfuscated} \\ \hline
    \textbf{Anderson et. al.\cite{Anderson12}} \ \   & 98.53\% \ \     & 91.87\%   \\
    \textbf{Our best model} \ \ & 99.26\%    \ \   & 93.18\%   \\ \hline
    \end{tabular}
    \end{table}
    
    Our proposed features perform slightly better in both obfuscated and unobfuscated malware detection, while maintaining lower false positive and false negative rates. The differences are slight at about $1\%$. However, the differences in the method are more apparent when looking at the power consumption for each feature extraction method.
  
    \begin{table}[ht]
    \caption{False Positive \& False Negative Rates. $X$\%/$Y$\% where $X$ is the false positive rate and $Y$ is the false negative rate where positive means malicious.}
    \label{tab:fpn}
    \vspace{1mm}
    \centering
    \begin{tabular}{c l l }
    FPR/FNR \ \ & \textbf{Unobfuscated} & \textbf{Obfuscated} \\ \hline
    \textbf{Anderson et. al.\cite{Anderson12}}     \ \      & 1.9\% / 1.2\%        & 1.5 \% / 13.1\%    \\
    \textbf{Our best model}  \ \       & 0.37 \% / 1.04\%       & 0.37\% / 11.98\%   \\ \hline
    \end{tabular}
    \end{table}
  
  \subsubsection*{Power consumption}
\begin{figure}[ht]
    \centering
    \begin{subfigure}[t]{0.48\textwidth}
        \centering
        \includegraphics[width=.77\linewidth]{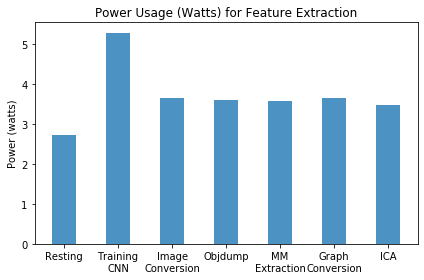}
        \caption{Watts}
    \label{fig:power1}
    \end{subfigure}
    ~
    \begin{subfigure}[t]{0.48\textwidth}
        \centering
        \includegraphics[width=.77\linewidth]{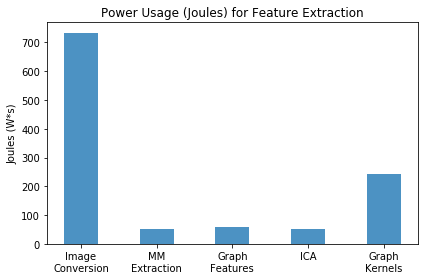}
        \caption{Joules}
    \label{fig:power2}
    \end{subfigure}
    \caption{Power usage during feature extraction measured in watts (a) and joules (b). Note that \textit{Graph Kernels} is the work of Anderson et. al. \cite{Anderson12}.}
    \label{fig:power}
  \end{figure}

  In Figures \ref{fig:power1} and \ref{fig:power2}, we show the power consumption, in watts and joules, for the feature extraction process used in our experiments. We present the power consumption needed to process 20 files. In addition to feature extraction, we included the power consumption of training a 2 hidden layer convolutional neural network for 50 epochs and a malware-to-image conversion \cite{Nataraj11} as baselines for comparison. All feature extraction processes take a similar amount of power in Watts. However, when we consider the time each process takes (measured in Joules), we can see that our proposed method is much more efficient. Even combining all three feature extraction processes consumes less energy than the graph kernel method proposed in \cite{Anderson12}. Note that while state-of-the-art models may use malware images and features extracted from such images, our results show that it is infeasible to do so on low powered devices. 
  
  Our experimental results show that context sensitive Markov matrices are an efficient representation of binaries from which features can be extracted (using independent component analysis and graph analysis) in resource constrained environments. With a maximum detection rate of 99.26\% and 93.18\% for unobfuscated and obfuscated malware, respectively, our proposed method slightly falls behind state of the art models that are run on powerful machines. However, our proposed features have significantly less computational and storage costs than these models, as well as compared to the related work. For example, even with a batch size of 1, we were not able to deploy any convolutional or recurrent neural network based models due to their storage cost at run-time. 

\section{Conclusion and future work} \label{conclusion}
Malware detection research has been transitioning towards detecting both unobfuscated and obfuscated samples. This is because, while new novel malware samples can be found, many are generated from existing strains. Many strides have been made in solving this problem. However, with increased use and deployment of IoT devices, a new attack vector has been introduced that has not yet been fully explored. 

In this paper, we have shown that we can maintain or have better detection accuracy than related malware detection features on IoT devices by using Markov matrices. With unobfuscated malware samples, we were able to achieve 99.26\% detection accuracy, which is less than 1\% behind state of the art models, such as DroidSieve for Android malware and MalConv for Windows executables \cite{Suarez17, raff2017}. Using Markov matrices also resulted in a $93.18\%$ detection rate of obfuscated malware. Furthermore, we showed that our proposed method of generating Markov matrices, as well as extracting features using ICA and simple graph algorithms, consumed less power compared to related work and state-of-the art models. 

There are several future research questions. We demonstrated that it is possible to transition machine learning based malware detection models to low resource systems while maintaining robustness against obfuscated malware samples. However, it would be interesting to expand upon rudimentary string matching techniques for generating control flow graph sensitive Markov matrices. An immediately apparent challenge would be limiting overhead. Our work suggests that research into lightweight dynamic analysis frameworks would be helpful in building malware detection into IoT devices. Additionally, this work considered only a subset of obfuscation techniques. Further research is needed for a solution against virtualization and packer-based obfuscation techniques. We hope that investigating these future questions will lead to secure low-powered IoT devices.

\bibliographystyle{IEEEtran}
\bibliography{mybib}

\end{document}